\documentclass[11pt]{article}
\usepackage{epsf}
\usepackage{amssymb}
\usepackage{amsmath}

\newcommand{\reseteqnum}{\setcounter{equation}{0}}

\title{
\hfill
\parbox{3cm}{\normalsize DPNU-00-21\\
{\tt  hep-lat/0005015}}\\
\vspace{0.5cm}
Gauge anomaly cancellations in \\ 
SU(2)$_L$$\times$ U(1)$_Y$ Electroweak theory on the lattice }
\author{
Yoshio Kikukawa\thanks{e-mail address:
kikukawa@eken.phys.nagoya-u.ac.jp} 
and 
Yoichi Nakayama\thanks{e-mail address:
yoichi@eken.phys.nagoya-u.ac.jp} 
\\
\\
{\normalsize\em Department of Physics, Nagoya University 
}\\
{\normalsize\em Nagoya 464-8602, Japan}
\\
\\
\date{\normalsize May, 2000}
}
\begin{document}

\maketitle

\begin{abstract}
We consider the cohomological classification of the 4+2-dimensional
topological field, which is proposed by L\"uscher,
for SU(2)$_L$ $\times$ U(1)$_Y$ electroweak theory.
The dependence on the admissible abelian gauge field of U(1)$_Y$
is determined through topological argument,
with SU(2)$_L$ gauge field fixed as background.
We then show the exact cancellation of the local gauge anomaly 
of the mixed type {SU(2)$_L$}$^2$ $\times$ U(1)$_Y$
at finite lattice spacing, as well as {U(1)$_Y$}$^3$,
using the pseudo reality of 
SU(2)$_L$ and the anomaly cancellation conditions in the electroweak 
theory 
given in terms of the hyper-charges of U(1)$_Y$.

\end{abstract}

% PACS 11.15.Ha, 12.38.Gc.

\newpage

\section{Introduction}
\label{sec:introduction}
\reseteqnum

The gauge interaction of the Weyl fermions now can be described 
in the framework of lattice gauge theory.
The clue to this development is the construction of gauge covariant 
and local Dirac operators \cite{overlap-D,fixed-point-D,
locality-of-overlap-D} which solve the Ginsparg-Wilson relation 
\cite{ginsparg-wilson-rel}. The Ginsparg-Wilson relation implies an
exact chiral symmetry for the Dirac fermion
\cite{exact-chiral-symmetry} and a 
gauge-field-dependent chiral projection to the Weyl degrees of freedom 
\cite{neidermayer-lat98,ginsparg-wilson-relation-and-overlap}. 

The functional measure for the Weyl fermion field is 
defined based on this chiral projection. It leads to a 
mathematically reasonable definition of the chiral determinant, 
which generically has the structure 
%known as the overlap formula \cite{overlap}. 
as an overlap of two vacua \cite{overlap}. 
It has been shown by L\"uscher in \cite{abelian-chiral-gauge-theory}
that for anomaly-free abelian chiral gauge theories, the functional 
measure for the Weyl fermion fields can be constructed so that the 
gauge invariance is maintained exactly on the lattice.
This issue of the gauge-invariant construction of 
the functional measure in non-abelian chiral theories has been 
related to the cohomological classification of a certain topological
field which is defined on the four-dimensional lattice
plus two continuum dimensions 
\cite{nonabelian-chiral-gauge-theory,lat99-luscher}.
It has been shown that in all orders in the 
lattice spacing $a$, the topological field has trivial 
cohomology for anomaly free theories.
This problem has also been examined by Suzuki 
from the point of view of the Wess-Zumino consistency condition 
and the BRST cohomology in four-dimensions \cite{BRST-cohomology-4D}. 
The gauge anomaly cancellation has been proved
for general gauge groups in all powers of gauge 
potential.\footnote{
The topological aspect of the non-abelian anomaly for Weyl fermions
defined based on the overlap formalism / the Ginsparg-Wilson relation 
has been examined by D.H.~Adams in close relation to the argument 
of L.~Alvarez-Gaum\'e  and P.~Ginsparg in the continuum theory 
\cite{Adams-AG,AG}.
The global SU(2) anomaly has been examined by H.~Neuberger and
O.~B\"ar and I.~Campos \cite{neuberger-su2,oliver-isabel} in detail. 
A lattice implementation of the $\eta$-invariant and
its relation to the effective action for chiral Ginsparg-Wilson
fermions has been examined by T.~Aoyama and Y.K. in
\cite{lattice-eta-invariant}.
Non-compact formulation of abelian chiral gauge theories
has been considered recently by 
Neuberger \cite{non-compact-abelian-chiral-gauge-theory}.
}

The above result for the abelian chiral gauge theories 
implies that U(1)$_Y$ hyper-charge chiral gauge theory now can 
be constructed on the lattice.
In this paper, we consider a first step towards the 
extension of this work to the case of 
the SU(2)$_L$$\times$ U(1)$_Y$ electroweak theory. 
We examine the exact cancellation of gauge anomalies 
in the SU(2)$_L$$\times$ U(1)$_Y$ electroweak theory,
through the cohomological classification 
\cite{topology-and-axial-anomaly,fujiwara-suzuki-wu}
of the 4+2-dimensional topological field proposed 
by L\"uscher \cite{nonabelian-chiral-gauge-theory}.
Here we will discuss the cancellation of the local anomaly 
in infinite volume lattice only 
and leave the issue related to possible global obstructions 
to the non-perturbative construction of the theory for 
future study.

The SU(2)$_L$$\times$U(1)$_Y$ electroweak theory contains 
the following fermions as the first generation:
\begin{equation}
  \left( \begin{array}{c} \nu_L(x) \\ 
                          e_L(x)     \end{array} \right)_{Y=-\frac{1}{2}},
\quad e_R(x)_{Y=-1} , 
\quad 
  \left( \begin{array}{c} u_{L \, i}(x) \\ 
                          d_{L \, i}(x) \end{array} \right)_{Y=\frac{1}{6}},
\quad 
  \begin{array}{c} u_{R \, i}(x)_{Y=+\frac{2}{3}} \\ 
                   d_{R \, i}(x)_{Y=-\frac{1}{3}} \end{array} ,
\end{equation}
where $i$ is the color index $(i=1,2,3)$.
The left-handed leptons and quarks are SU(2)$_L$ doublets. 
The right-handed fermions are SU(2)$_L$ singlet. Taking into 
account of the color degrees of freedom, there are four doublets.
The hyper-charge $Y$, which is related to electromagnetic charge $Q$
by the Gell-Mann-Nishijima relation,
\begin{equation}
  Q= I_3 + Y ,
\end{equation}
are assigned as shown above.

There are two types of gauge anomalies in the electroweak theory.
The first one is the gauge anomaly associated with the abelian 
U(1)$_Y$ gauge group. 
The second one is the gauge anomaly of the mixed type among
SU(2)$_L$ and U(1)$_Y$ gauge groups.  In the continuum theory,
these gauge anomalies are generated from the following diagrams:
\begin{figure}[h]
  \begin{center}
\leavevmode
\epsfxsize=80mm
\epsfbox{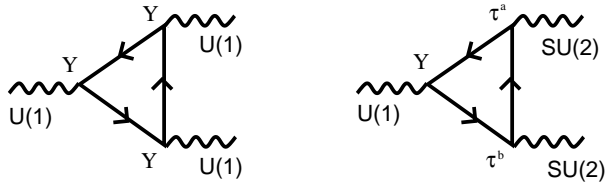}
\caption{Gauge Anomalies in the SU(2)$_L$$\times$U(1)$_Y$ electroweak theory}
\label{fig:gauge-anomaly}
  \end{center}
\end{figure}
\newline
Then the conditions for the gauge anomaly cancellation in
the electroweak theory are given in terms of the hypercharges as
\begin{equation}
 \label{anomaly-cancellation-YYY}
 \sum_L Y^3 - \sum_R Y^3 =0 ,
\end{equation}
and
\begin{equation}
\sum_{\mbox{doublet(L)}} Y =0. 
 \label{anomaly-cancellation-Y}
\end{equation}
We can see that the assignment of the hyper-charges shown above
indeed satisfies these conditions and one more condition as 
\begin{equation}
\sum_{\mbox{singlet(R)}} Y =0 .
\end{equation}

In order to show the exact cancellations of the (local) gauge anomalies
at finite lattice spacing,
we consider the cohomological classification 
of the 4+2-dimensional
topological field for SU(2)$_L$ $\times$ U(1)$_Y$ electroweak theory.
Our approach is then to determine the dependence on the admissible 
abelian gauge field of U(1)$_Y$ through topological argument, 
{\it with SU(2)$_L$ gauge field fixed as background}
(cf. \cite{topology-and-axial-anomaly,fujiwara-suzuki-wu}).
Although it does not determine the explicit dependence on 
SU(2)$_L$ gauge field, it turns out to be sufficient to show
the exact cancellations of the gauge anomalies:
we can show the cancellation of the gauge anomaly 
of the mixed type {SU(2)$_L$}$^2$ $\times$ U(1)$_Y$
at finite lattice spacing, as well as {U(1)$_Y$}$^3$,
using the pseudo reality of 
SU(2)$_L$ and the anomaly cancellation conditions in the electroweak 
theory given in terms of the hyper-charges of U(1)$_Y$.

This paper is organized as follows. 
In section~\ref{sec:q-for-electroweak-theory},
we introduce the 4+2 dimensional topological field for
the electroweak theory and discuss its specific features for 
SU(2)$_L$ $\times$ U(1)$_Y$ gauge groups.
In section~\ref{sec:cohomological-classification},
we formulate the Poincar\'e lemma in 4+2 dimensions and 
determine the dependence on the admissible U(1)$_Y$ field 
at finite lattice spacing.
In section~\ref{sec:exact-cancellations}, we show
the exact cancellations of gauge anomalies 
of the mixed type {SU(2)$_L$}$^2$ $\times$ U(1)$_Y$
as well as {U(1)$_Y$}$^3$,
using the pseudo reality of 
SU(2)$_L$ and the anomaly cancellation conditions in the electroweak 
theory given in terms of the hyper-charges of U(1)$_Y$.
In section~\ref{sec:discussion}, we give some discussions.

\section{4+2 dimensional topological field for 
Electroweak theory on the lattice}
\label{sec:q-for-electroweak-theory}
\reseteqnum

\subsection{Weyl fermions on the lattice}
Let us consider lattice Dirac fermion which
is described by a gauge-covariant and local lattice 
Dirac operator which satisfies the Ginsparg-Wilson relation.
\begin{equation}
  D \gamma_5 + \gamma_5 D = a D \gamma_5 D .
\end{equation}
The action of the Dirac fermion is written as
\begin{equation}
  S = a^4 \sum_x \bar \psi(x) D \psi(x) .
\end{equation}

In the case of Neuberger's Dirac operator
\begin{equation}
  D= \frac{1}{a}\left( 1 + \gamma_5 \frac{H}{\sqrt{H^2}} \right), 
\end{equation}
where $H$ is defined by the hermitian Wilson-Dirac operator
\begin{equation}
  H = \gamma_5 \left( \sum_\mu \left\{
 \frac{1}{2} \gamma_\mu \left( \nabla_\mu-\nabla_\mu^\dagger \right)
+\frac{a}{2} \nabla_\mu \nabla_\mu^\dagger \right\} - \frac{m_0}{a} \right),
\end{equation}
locality of the action has been proved rigorously for gauge fields with
bounded field strength \cite{locality-of-overlap-D,bound-neuberger}.
\begin{equation}
  \| 1 - U_{\mu\nu}(x) \| \le \epsilon, \quad
1-6(2+\sqrt{2}) \epsilon  > | 1-m_0|^2 .
\end{equation}
This proof has been extended to the case where $H$ is 
defined by the transfer matrix of the five-dimensional Wilson 
fermion \cite{locality-kikukawa}.

The action is invariant under the transformation which
can be regarded as the chiral transformation on the lattice:
\begin{equation}
\delta   \psi(x)  = \gamma_5\left(1-aD\right) \psi(x) , \qquad
\delta   \bar \psi(x)  = \bar \psi(x) \gamma_5 .
\end{equation}

By virtue of this exact chiral symmetry, we can define
left-handed Weyl fermion  on the lattice by the following projections:
\begin{equation}
 \hat P_- \psi_L(x) = \psi_L(x), \quad \bar \psi_L(x) P_+ = \bar \psi_L(x).
\end{equation}
$\hat P_-$ is the chiral projector defined as 
\begin{equation}
 \hat P_- = \frac{1-\hat\gamma_5}{2}, \qquad
 \hat \gamma_5 = \gamma_5 (1-aD).
\end{equation}
$P_+$ is the usual chiral projector defined with $\gamma_5$.
The right-handed Weyl fermions can be defined in the similar manner.

The functional measure for the Weyl fermion can be defined as follows:
we first introduce chiral bases $\left\{ v_j(x) \right\}$ and
$\left\{ \bar v_k(x) \right\}$ as 
\begin{equation}
\hat P_- v_j(x) = v_j(x), \qquad 
\bar v_k(x) P_+ = \bar v_k(x),
\end{equation}
and expand the Weyl fermion fields in terms of the chiral bases 
with the coefficients which generate the Grassmann algebra,
\begin{equation}
  \psi(x) = \sum_j v_j(x) c_j , \qquad 
  \bar \psi(x) = \sum_k \bar c_k \bar v_k(x).
\end{equation}
Then the functional measure of the Weyl fermion can be defined as
\begin{equation}
 \prod_x d\psi_L(x) d\bar\psi_L(x) 
=\prod_j d c_j \prod_k d \bar c_k .
\end{equation}
Given the definition for the functional measure of the Weyl fermion, 
the chiral determinant is evaluated as 
\begin{eqnarray}
  Z_W&=& \int  \prod_x d\psi_L(x) d\bar\psi_L(x)  
\exp\left(- a^4 \sum_x \bar \psi_L(x) D \psi_L(x) \right) \\
     &=& \det M_{kj} , 
\end{eqnarray}
where
\begin{equation}
M_{kj}= a^4 \sum_x \bar v_k(x) D v_j(x) = \left( \bar v_k D v_j \right).
\end{equation}

\subsection{4+2 dimensional topological field}

The question is how to construct the functional measure 
for the Weyl fermions so that gauge invariance is maintained
at finite lattice spacing. 
As shown by L\"uscher \cite{nonabelian-chiral-gauge-theory}, 
this question can be formulated as the cohomological problem of 
a certain %4+2 dimensional
topological field which is defined on the four-dimensional lattice
plus two continuum dimensions.

The 4+2 dimensional topological field is introduced as follows.
We consider lattice gauge fields
\begin{equation}
  U_\mu(z) \in G, \quad z=(x_\mu,t,s), \quad \mu=1,2,3,4
\end{equation}
which depend on two additional real coordinates  $t$ and $s$.
We also introduce gauge potentials $A_t(z)$ and $A_s(z)$ along
these directions and define the associated field tensor by
\begin{equation}
F_{ts}(z)=
\partial_t A_s(z)-  \partial_s A_t(z) + i \left[ A_t(z), A_s(z) \right].
\end{equation}
The covariant derivative in these directions is defined as 
\begin{equation}
D_r^A U_\mu(z) = \partial_r U_\mu(z)
+ i \left[ A_r(z) U_\mu(z)- U_\mu(z) A_r(z+a \hat\mu)
    \right] , \quad r=t,s
\end{equation}
which transforms in the same way as $U_\mu(z)$
under gauge transformations in 4+2 dimensions.
Then we consider the following 4+2 dimensional field which 
is gauge invariant and local:
\begin{eqnarray}
&& q(z) = 
 -i\, \mbox{tr}\, \left\{\left[\frac14 \hat\gamma_5 \left[D_t^A\hat P_- ,
 D_s^A\hat P_- \right]+\frac14 \left[D_t^A\hat P_-,D_s^A \hat P_-\right]
 \hat\gamma_5
 \right.\right.  \nonumber\\ 
&& \qquad\qquad\qquad
\left.\left.
 +\frac{i}{2} R(F_{ts})\hat\gamma_5\right](x,x)\right\}. 
\label{definition-of-q} 
\nonumber\\
\end{eqnarray}
The trace is taken over the Dirac and flavor indices only. 

By noting 
\begin{equation}
a^4 \sum_x q(x)=
i {\rm Tr}\left\{ 
\hat P_- \left[ \partial_t \hat P_-, \partial_s \hat P_-\right]
-\frac{i}{2} \partial_t \left[ R(A_s) \hat \gamma_5 \right]
+\frac{i}{2} \partial_s \left[ R(A_t) \hat \gamma_5 \right]
%-\frac{1}{2}R(F_{ts}) \hat \gamma_5 
\right\}
\end{equation}
and making use of the identity
\begin{equation}
  {\rm Tr}\left\{ \delta_1 \hat P_- \delta_2 \hat P_- \delta_3 \hat P_-
  \right\} = 0 ,
\end{equation}
we can show that this 4+2 dimensional field satisfies 
\begin{equation}
 a^4\sum_{x}\int dt ds \delta q(z) =0
\end{equation}
for all local variations of the link variables $U_\mu(z)$ and
the potential $A_r(z)$, i.e. it is a topological field.

It has been shown by L\"uscher \cite{nonabelian-chiral-gauge-theory} 
that if this topological field is in the trivial cohomology class, 
i.e. it is equal to the divergence of a gauge-invariant
local current,
\begin{equation}
 q(z)=\partial_\mu^* k_\mu(z) +\partial_t k_s(z)-\partial_s k_t(z),
 \label{q-is-total-divergence}
\end{equation}
then using $k_r(z)$ it is possible to construct a 
gauge-covariant local current $j_\mu^a(x)$ 
which satisfies the integrability condition in differential form
and the anomalous conservation law. 

%with desired properties,
%from which the gauge-invariant functional measure for the Weyl
%fermions can be reconstructed.

\subsection{4+2 dimensional topological field
for SU(2)$_L$$\times$U(1)$_Y$ electroweak theory}

In order to construct the 4+2 dimensional topological field
for SU(2)$_L$$\times$U(1)$_Y$ electroweak theory, 
we consider lattice SU(2)$_L$ and U(1)$_Y$ gauge fields,
\begin{equation}
  U^{(1)}_\mu(z) \in {\rm U(1)}, 
\qquad U^{(2)}_\mu(z) \in {\rm SU(2)},
\end{equation}
which satisfy the admissibility conditions with sufficiently small constants 
$\epsilon^{(2)}$ and $\epsilon^{(1)}$:
\begin{equation}
  \| 1- U^{(2)}_{\mu\nu}(x) \| < \epsilon^{(2)} , \qquad
  \| 1- U^{(1)}_{\mu\nu}(x) \| < \epsilon^{(1)} .
\end{equation}

When $ \epsilon^{(1)} < 1/ 6 Y  \times \pi/3 $, 
the admissible abelian lattice gauge fields can be 
expressed in terms of vector potentials 
\begin{equation}
 U^{(1)}_\mu(x) = \exp \left(iA_\mu(x)\right)
\end{equation}
which has the following properties:
\begin{equation}
F_{\mu\nu}(x) \equiv \frac{1}{i} \ln U_{\mu\nu}(x) 
=\partial_\mu A_\nu (x)-\partial_\nu A_\mu (x) , \quad
|A_\mu(x)| \leq \pi (1+8\|x\|). 
\end{equation}
This representation of the link variable is unique up to 
the gauge transformation with the parameter 
$\omega(x)$ which takes values in integer multiple of $2\pi$.
\begin{equation}
 \widetilde A_\mu(x)=A_\mu(x) +\partial_\mu \omega(x).
\end{equation}

We also introduce gauge potentials along the two additional dimensions
$A_r(z), r=t,s$ for U(1)$_Y$ and $B_r(z), r=t,s$ for SU(2)$_L$
and denote the 4+2 dimensional gauge fields as follows:
\begin{equation}
  A_\mu(z) = \left( A_k(z), A_t(z), A_s(z) \right) ,
\quad 
  U_\mu(z) = \left( U^{(2)}_k(z), i B_t(z), i B_s(z) \right) ,
\end{equation}
where $\mu=1,\cdots,6$ and we use the Latin index $i=1,2,3,4$ for
four-dimensional lattice here after.
Then the 4+2 dimensional topological field $q(z)$ for 
SU(2)$_L$$\times$U(1)$_Y$ electroweak theory can be regarded as 
a gauge-invariant local functional of the 
4+2 dimensional gauge field variables $A_\mu(z)$ and $U_\mu(z)$:
\begin{equation}
  q(z)= q\left(z; A_\mu(z) , U_\mu(z) \right)
\end{equation}

It follows from the charge conjugation property of 
the lattice Dirac operator that 
$q(z)$ changes sign under complex-conjugation of the representations 
of the gauge fields
\begin{equation}
 q\left(z; A_\mu(z) , U_\mu(z) \right)
= - q\left(z; -A_\mu(z), U^\ast_\mu(z) \right)
\end{equation}
Since SU(2)$_L$ is pseudo real, 
there exists a unitary transformation $S$ such that
\begin{equation}
\label{pseudo-reality-of-su2}
 S U^\ast_\mu S^{-1} = U_\mu .
\end{equation}
Then we obtain 
\begin{equation}
 q\left(z; A_\mu(z),U_\mu(z)  \right)
= - q\left(z; -A_\mu(z),U_\mu(z) \right).
\end{equation}
We note also that $q(z)$ vanishes identically when
the U(1)$_Y$ gauge fields are switched off:
\begin{equation}
 q\left(z; 0, U_\mu(z) \right)  = 0.
\end{equation}

\section{Cohomological classification of the topological field in 4+2
 dimensions}
\label{sec:cohomological-classification}
\reseteqnum

\subsection{Analysis of the 4+2 dimensional topological field}

In this section, 
%let us examine the 4+2 dimensional
%topological field $q(x,t,s)$ when varying U(1) gauge field.
we will formulate the Poincar\'e lemma in 4+2 dimensions and 
examine the dependence of $q(x,t,s)$ on the admissible U(1)$_Y$ field 
through topological argument. In the course of the analysis, 
SU(2)$_L$ gauge field is fixed as background.

Since $q(x,t,s)$  smoothly depends on 4+2 dimensional U(1) gauge potential
$A_\mu(x,t,s)$ and its differentials, 
the variation of the topological field can be expressed as
\begin{equation}
\label{eq:q-variation}
 \delta q(x,t,s)
 = 
%\sum_{m \le 1,n \le 1}\sum_{y}
\sum_{m,n=0,1}\sum_{y}
 \frac{\partial q(x,t,s)}{\partial [\partial_s^m\partial_t^n A_\mu(y,t,s)]}
 \partial_s^m\partial_t^n \delta A_\mu (y,t,s).
\end{equation}
By definition, $q(x,t,s)$ contains at most the first-order
differentials of the vector potential $A_\mu(x,t,s)$ in the continuous
coordinates. Therefore, we may restrict the sum over $m,n$ to $0,1$.

If we define the differential operator in 4+2 dimensions in the above
expression as 
\begin{equation}
\label{eq:operator-L}
 L_\mu(x,y,t,s)= \sum_{m,n=0,1}
 \frac{\partial q(x,t,s)}{\partial[\partial_s^m\partial_t^n A_\mu(y,t,s)]}
 \partial_s^m\partial_t^n ,
\end{equation}
then the topological property and the gauge invariance 
of the 4+2 dimensional field lead to the following conditions for
the differential operator $L_\mu$.
\begin{equation}
\label{eq:q-topological-property}
\int dt ds \, a^4 \sum_x \, \sum_y L_\mu(x,y,t,s) \delta A_\mu(y,t,s) = 0 ,
\end{equation}
and
\begin{equation}
\label{eq:q-gauge-invariance}
L_\mu(x,y,t,s)  \partial_\mu   = 0.  
\end{equation}
The topological field $q(x,t,s)$ itself can be expressed with this 
operator as 
\begin{equation}
\label{eq:q-in-L}
q(x,t,s) 
= \alpha(x,t,s) + \sum_{y} \left. 
\int_{0}^{1} d u \, L_\mu (x,y,t,s) \right|_{A\rightarrow u A} 
 \, A_\mu (y,t,s),
\end{equation}
where $\alpha(x,t,s)$ is the part which does not depend on the abelian
gauge field $A_\mu(x,t,s)$.
Then the problem reduces to examine the cohomological properties of the
differential operator $L_\mu$. 
In order to examine such an operator and determine the form 
of the topological field $q(x,t,s)$, 
we will next formulate the Poincar\'e lemma which is applicable to 
the differential operators in 4+2 dimension.
This is the extension of the Poincar\'e lemma on the lattice given 
in \cite{topology-and-axial-anomaly}
along the line of the analysis in the continuum theory of
\cite{luscher-unpublished}. 

\subsection{Poincar\'e lemma in 4+2 dimensions}

We first introduce a Grassmann algebra with basis element
\begin{equation}
d x_1, d x_2, d x_3, d x_4, d x_5 =dt, d x_6=ds  
\end{equation}
and denote them by $dx_\mu (\mu=1,\cdots 6)$ collectively.
A $k$-form ($0 \le k \le 6$) in 4+2 dimensions is then defined 
as:
\begin{equation}
  f(z)= \frac{1}{k!} f_{\mu_1 \cdots \mu_k}(z) 
dx_{\mu_1} \cdots dx_{\mu_k} \, \in \Omega_k , \quad z=(x_i,t,s).
\end{equation}
We assume that $f_{\mu_1 \cdots \mu_k}(z)$ is a smooth function in
the continuous coordinates $t,s$ and is locally supported in the lattice
coordinate $x_i$.

The exterior differential operator, 
which is a map $\Omega_k  \rightarrow \Omega_{k+1}$, is defined by
$d = \sum_{i=1}^4 d x_i \partial_i +  \sum_{r=t,s} d r\partial_r$
where $\partial_i$ is forward difference operator on the
four-dimensional lattice.
It acts on the $k$-forms according to 
\begin{equation}
d f(z) = \frac{1}{k!} \partial_\mu f_{\mu_1 \cdots \mu_k}(z) 
\, d x_\mu dx_{\mu_1} \cdots dx_{\mu_k}.
\end{equation}
The associated divergence operator 
$d^\ast : \Omega_k \rightarrow \Omega_{k-1}$ is defined as
\begin{equation}
 d^\ast f(z) = \frac{1}{(k-1)!} 
\partial_\mu^* f_{\mu \mu_2 \cdots \mu_k}(z) \, dx_{\mu_2} \cdots dx_{\mu_k},
\end{equation}
where $\partial_i^*$ is backward difference operator.

Now we consider a class of differential operators $L$ 
which is a map $L : \Omega_l  \rightarrow \Omega_k$ such that
\begin{equation}
L \, f(x,t,s) 
= \frac{1}{k! l!} d x_{\mu_1} \cdots d x_{\mu_k} 
\sum_{y,n,m}  
L^{n,m}_{\mu_1 \cdots \mu_k;\nu_1 \cdots \nu_l}(x;y,t,s) \, 
\partial_t^n \partial_s^m 
\, f_{\nu_1 \cdots \nu_l}(y,t,s),
\end{equation}
where $L^{n,m}_{\mu_1 \cdots \mu_k;\nu_1 \cdots \nu_l}(x;y,t,s)$ 
is a local function in $x(y),t,s$ which is exponentially decaying 
in $x(y)$ with respect to the reference point $y(x)$.
We assume $m,n =0,1$ in the following discussions.
When $L$ is not a differential operator, i.e. $n=m=0$ and
is proportional to the Kronecker delta $\delta_{x,y}$,
we refer such an operator as zero degree.
\begin{equation}
L^0 f(x,t,s) 
= \frac{1}{k! l!} d x_{\mu_1} \cdots d x_{\mu_k} 
L^0_{ \mu_1 \cdots \mu_k;\nu_1 \cdots \nu_l}(x,t,s) \, 
\, f_{\nu_1 \cdots \nu_l}(x,t,s).
\end{equation}

We can show the following Poincar\'e lemma for these 
differential operators in 4+2 dimensions.

\noindent
{\bf Lemma} (Poincar\'e lemma)
If $L : \Omega_l \rightarrow \Omega_k$ is 
any differential operator satisfying 
\begin{equation}
\label{eq:if-closed}
 d L(x,y,t,s)=0 ,
\end{equation}
then there exists a differential operator $M(x,y,t,s)$ 
and an operator of zero degree $L^0$
such that
\begin{equation}
 L(x,y,t,s)= \delta_{k,6} \delta_{x,y}L^0(x,t,s)+ d M(x,y,t,s),
\end{equation}
Note that the product of $d$ and $M$ is a product of operators.

The proof of the lemma is given as follows.
We first note the fact that for any differential operator $L$ in concern
and any fixed continuous dimension $r$, 
there is a unique decomposition of the operator into two 
operators so that 
\begin{equation}
\label{eq:decomposition-degree-zero}
    L = S + \partial_r R,
\end{equation}
where $S$ is zero degree with respect to $\partial_r$. 
This is because $L$ is at most a first-order differential operator 
in terms of $\partial_r$ assuming the form
\begin{equation}
  L = L_0 + L_1 \partial_r
\end{equation}
with $L_0$ and $L_1$ zero degree with respect to $\partial_r$
and it can be rewritten uniquely into 
\begin{equation}
L = \left( L_0-  [ \partial_r L_1 ] \right) + \partial_r L_1,
\end{equation}
where the bracket $[ \partial_r L_1 ]$ stands for the fact 
that $\partial_r$ in it acts only on $ L_1 $.
When $L$ is gauge invariant, by acting it on a constant, we 
infer that $L_0$ is also gauge invariant. 
Then it also follows that $L_1$ is gauge invariant. 
Therefore, in the decomposition 
of Eq.~(\ref{eq:decomposition-degree-zero}), both $S$ and $R$
are gauge invariant.

Then we can decompose any differential operator 
$L : \Omega_l \rightarrow \Omega_k$ into the sequence:
\begin{equation}
 L=(L_5+\partial_s R_5)ds +Z_6,
\end{equation}
and
\begin{equation}
 L_5=(L_4+\partial_t R_4) dt +Z_5,
\end{equation}
where $L_5$ is degree zero with respect to $\partial_s$ and
$L_4$ is degree zero with respect to both $\partial_s$ and 
$\partial_t$. 

From the condition Eq.~(\ref{eq:if-closed}), 
\begin{equation}
  d L = \tilde d \left( L_5 + \partial_s R_5 \right) ds 
+ \left( 
\tilde d + ds \partial_s \right) Z_6 = 0,
\end{equation}
where $\tilde d= \sum_{i=1}^4 d x_i \partial_i + d t \partial_t$.
Then the coefficient of $ds$ must vanish
\begin{equation}
\tilde d L_5 + \partial_s \left( \tilde d R_5+ (-)^k Z_6 \right) = 0.
\end{equation}
Since this can be regarded as the unique decomposition of zero 
operator with respect to $\partial_s$, we have
\begin{eqnarray}
\label{eq:if-closed-L5}
&&  \tilde d L_5 = 0 , \\
&&  \tilde d R_5+ (-)^k Z_6 = 0.
\end{eqnarray}
In a similar manner, from the condition Eq.~(\ref{eq:if-closed-L5}), 
we obtain
\begin{eqnarray}
\label{eq:if-closed-L4}
&&  \bar d L_4 = 0 , \\
&&  \bar d R_4+ (-)^{k-1} Z_5 = 0,
\end{eqnarray}
where $\bar d =\sum_{i=1}^{4}dx_i \partial_i$.

Combing these results, we have
\begin{equation}
 L = L_4 dt ds + d K, 
\end{equation}
where
\begin{equation}
  K = (-)^{k-1} \, R_5 + (-)^{k-2} \, R_4 \, ds.
\end{equation}
$L_4$ is zero degree with respect to both $\partial_s$ and
$\partial_t$ and satisfies $\bar d L_4 =0$. 

Now we recall the Poincar\'e lemma on four-dimensional lattice given in
\cite{topology-and-axial-anomaly}. From the condition $\bar d L_4 =0$, 
we have
\begin{equation}
  L_4(x;y,t,s) = \delta_{k-2,4} \delta_{x,y} L_4^0(x,t,s) 
+ \bar d T(x;y,t,s)
\end{equation}
where
\begin{equation}
 L_4^0(x,t,s) = \sum_y L_4(y;x,t,s).
\end{equation}
Then we can write $L$
\begin{equation}
 L(x,y,t,s)= \delta_{k,6} \delta_{x,y}L^0(x,t,s)+ d M(x,y,t,s),
\end{equation}
where
\begin{equation}
  M= T dt ds + K, \qquad   L^0 = L_4^0 dt ds .
\end{equation}

It is clear from the above construction that 
$L^0$ and $M$ possesses the same locality 
and gauge-transformation properties as $L$.
When $L$ is gauge invariant under SU(2)$_L$ and U(1)$_Y$ 
gauge transformations,  $L_0$ and $M$ are also gauge invariant.

For the operators $L$ which satisfy $ d^\ast L = 0$, 
we can show the following form of the Poincar\'e lemma: 
If $L : \Omega_l \rightarrow \Omega_k$ is 
any differential operator satisfying 
\begin{equation}
\label{eq:if-closed-d-ast}
 d^\ast L(x,y,t,s)=0 ,
\end{equation}
then there exists a differential operator $M(x,y,t)$ 
and an operator of zero degree $L^0$
such that
\begin{equation}
 L(x,y,t,s)= \delta_{k,0} \delta_{x,y}L^0(x,t,s)+ d^\ast M(x,y,t,s).
\end{equation}
For the operators $L$ which satisfy $ L d = 0$
and $L d^\ast =0$, the lemma can be derived in similar manners.

\subsection{Structure of the 4+2 dimensional topological field}

By using the Poincar\'e lemma in 4+2 dimensions, we next determine the 
dependence of $q(z)$ on the admissible abelian gauge field of U(1)$_L$. 
We will show the following lemma:

\vspace{1em}
\noindent
{\bf Lemma} % (Chern class in 4+2 dim. U(1) gauge theory)
The 4+2 dimensional topological field $q(z)$ 
for the SU(2)$_L$ $\times$ U(1)$_Y$ electroweak theory 
is written in the following form.
\begin{eqnarray}
\label{eq:classification-q}
q(z)&=&
  \alpha(z) 
+ \beta_{\mu\nu}(z-\hat\mu-\hat\nu) 
  {F}_{\mu\nu}(z-\hat\mu-\hat\nu) 
\nonumber\\ 
&+&  
\gamma_{\mu\nu\rho\sigma}(z-\hat\mu-\hat\nu-\hat\rho-\hat\sigma) 
{F}_{\rho\sigma}(z-\hat\mu-\hat\nu-\hat\rho-\hat\sigma) 
{F}_{\mu\nu} (z-\hat\mu-\hat\nu)
\nonumber\\
&+& \delta \, \epsilon_{\mu\nu\rho\sigma\lambda\tau}
{F}_{\lambda\tau}
(z-\hat\mu-\hat\nu-\hat\rho-\hat\sigma-\hat\lambda-\hat\tau) 
  \times \nonumber\\
&& \qquad \qquad
{F}_{\rho\sigma} (z-\hat\mu-\hat\nu-\hat\rho-\hat\sigma) \,
{F}_{\mu\nu}(z-\hat\mu-\hat\nu) \,
\nonumber\\
&+& \partial^*_\mu k_\mu(z)
\end{eqnarray}
where
${F}_{\mu\nu}= \partial_\mu A_\nu - \partial_\nu A_\mu$
is the 4+2 dimensional field strength of U(1)$_Y$ gauge potential.
$k_\mu(z)$ is a local current which is gauge invariant 
under SU(2)$_L$ and U(1)$_Y$ gauge transformations.
$\delta$ is a constant.
$\alpha(z)$, $\beta_{\mu\nu}(z)$ and
$\gamma_{\mu\nu\rho\sigma}(z)$ are
certain gauge invariant local functions which may depend on 
the SU(2)$_L$ gauge field and satisfy
\begin{equation}
  \partial_\mu^* \beta_{\mu\nu}(z) = 0 , \quad
  \partial_\mu^* \gamma_{\mu\nu\rho\sigma}(z)= 0.
\end{equation}
Note also that $\hat 5 = \hat 6 =0$ and $z=(x_i,t,s)$.

\vspace{1em}
The proof consists of three steps.

\subsubsection{Step one}

We first consider the differential operator 
\begin{equation}
 L_\mu(x,y,t,s)= \sum_{m,n=0,1}
 \frac{\partial q(x,s,t)}{\partial[\partial_s^m\partial_t^n A_\mu(y,s,t)]}
 \partial_s^m\partial_t^n .
\end{equation}
We may regard this operator as a map $L_\mu : \Omega_1 \rightarrow \Omega_0$.
Then using the Poincar\'e lemma, we can write as 
\begin{equation}
\label{eq:L=dK}
 L_\mu(x,y,t,s) = \delta_{x,y}L^0_\mu (x,t,s) + 
\partial^*_\nu K_{\nu;\mu},
\end{equation}
where $L^0$ is zero degree and $K_{\nu;\lambda}$ is a map
$K_{\nu;\lambda}: \Omega_1 \rightarrow \Omega_1$.
The topological property of the 4+2 dimensional field implies
\begin{equation}
 0 = \int dt ds \, a^4 \sum_{x} \delta q(x,t,s)
 = \int dt ds \,  a^4 \sum_{x} L^0_{\mu}(x,t,s) \delta A_\mu (x,t,s) .
\end{equation}
Since $L^0_{\mu}(x,t,s)$ is zero degree, it must vanish identically,
\begin{equation}
  L^0_\mu(x,t,s) = 0.
\end{equation}

On the other hand, the gauge invariance of the 4+2 dimensional 
field implies 
\begin{equation}
L_\mu(x,y,t,s)  \partial_\mu   = 0
\end{equation}
or
\begin{equation}
\partial^*_\nu K_{\nu; \mu}(x,y,t,s) \partial_\mu  = 0.
\end{equation}
Then, using the Poincar\'e lemma, we obtain
\begin{equation}
 K_{\nu;\mu}(x,y,t,s) \partial_\mu 
= \partial^*_\lambda H_{\lambda\nu} (x,y,t,s) ,
\label{Kd=dH}
\end{equation}
where $H_{\lambda\nu}: \Omega_0 \rightarrow \Omega_2$.
Using the Poincar\'e lemma again, it can be cast 
into the form
\begin{equation}
 H_{\lambda\nu}(x,y,t,s) = \delta_{x,y}H^0_{\lambda \nu}(x,t,s) 
 + R_{\lambda \nu;\rho}(x,y,t,s)\partial_\rho ,
\end{equation}
where $R_{\lambda\nu;\rho}: \Omega_1 \rightarrow \Omega_2$.
We can eliminate $R_{\lambda \nu;\rho}$ term 
in the above expression by redefining $K_{\nu;\mu}$ so that
\begin{equation}
  K_{\nu;\mu} \longrightarrow 
K_{\nu;\mu} + \partial^*_\lambda R_{\lambda \nu;\mu}(x,y,t,s),
\end{equation}
which does not affect the relation 
$L_\mu=\partial^*_\nu K_{\nu;\mu}$. 
%(Since $L=d^*K$, we can redefine $K\rightarrow K + d^* R$.)
Thus we obtain
\begin{equation}
 H_{\lambda \nu}(x,y,t,s) = \delta_{x,y}H^0_{\lambda \nu}(x,t,s).
\end{equation}
We substitute this result into Eq.~(\ref{Kd=dH}) and make it act 
on a constant, we obtain
\begin{equation}
 [\partial_\lambda^* H^0_{\lambda \nu}(x,t,s)]=0,
\end{equation}
where the square bracket means that $\partial_\lambda^*$ acts on the 
function $H^0_{\lambda \nu}$ rather than a product of the differential
operators. Then it follows that\footnote{
  \begin{eqnarray}
%&&
%\partial_\lambda^* 
%\left[
%\sum_y \delta_{x,y} 
%H^0_{\lambda \nu}(x,t,s)f(y,t,s) \right]
%\nonumber\\
&&
\partial_\lambda^* \left[ H^0_{\lambda \nu}(x,t,s)f(x,t,s)\right]
\nonumber\\
&&= 
\left[ \partial_\lambda^* H^0_{\lambda \nu}(x,t,s) \right] f(x,t,s)
+ H^0_{\lambda \nu}(x-\hat \lambda,t,s)  
\left[ \partial_\lambda^* f(x,t,s) \right] 
\nonumber\\
&&= 
\left[ \partial_\lambda^* H^0_{\lambda \nu}(x,t,s) \right] f(x,t,s)
+ H^0_{\lambda \nu}(x-\hat \lambda,t,s)  
\left[ \partial_\lambda f(x-\hat \lambda,t,s) \right] 
.
\nonumber
%\nonumber\\
%&&= 
% \delta_{x-\hat \lambda,y}\, 
%H^0_{\lambda \nu}(x-\hat \lambda,t,s) \partial_\lambda.
  \end{eqnarray}
}
\begin{equation}
 \partial_\lambda^* \delta_{x,y} H^0_{\lambda \nu}(x,t,s)
 = %\delta_{x,y}\left[\partial_\nu^* H_{0,\nu\mu}(x,t)\right]+
 \delta_{x-\hat \lambda,y}\, 
H^0_{\lambda \nu}(x-\hat \lambda,t,s) \partial_\lambda.
\end{equation}
Then using the Poincar\'e lemma we obtain 
\begin{equation}
 K_{\nu;\mu}(x,y,t,s) 
=  \delta_{x-\hat \mu,y}\, H^0_{\mu \nu}(x-\hat \mu,t,s) 
  + \omega_{\nu;\mu\rho}(x,y,t,s) \partial_\rho.
\end{equation}

We now substitute this result into Eq.~(\ref{eq:L=dK}) and obtain
\begin{equation}
 L_\mu(x,y,t,s) 
=\delta_{x-\hat\nu-\hat\mu,y}
 H^0_{\mu\nu}(x-\hat\nu-\hat\mu,t,s)\partial_\nu 
+\partial_\nu^* \omega_{\nu;\mu\rho}(x,y,t,s)\partial_\rho .
\end{equation}
Then the 4+2 dimensional topological field is given as follows.
\begin{eqnarray}
q(x,t,s) 
&=& \alpha(x,t,s) + \int_{0}^{1} du \left. \sum_{y}
 L_\mu (x,y,t,s)\right|_{A\rightarrow u A}
 \hspace{-8mm}A_\mu (y,t,s)\nonumber\\
&=& 
\alpha(x,t,s) +
\left. \int_{0}^{1} du  \left[ H^0_{\mu\nu}(x-\hat\nu-\hat\mu,t,s)
 \right]\right|_{A\rightarrow uA}\hspace{-8mm}\partial_\nu
  A_\mu(x-\hat\nu-\hat\mu,t,s) 
\nonumber\\
&& +  \partial_\nu^* 
\left( 
\left. \sum_{y}\int_{0}^{1} du \, 
 \omega_{\nu,\mu\rho}(x,y,t,s)\right|_{A\rightarrow uA}
 \hspace{-8mm}\partial_\rho A_\mu(y,t,s) \right).
\end{eqnarray}
Setting 
\begin{eqnarray}
\phi_{\mu\nu}(x,t,s)
&=& - \frac{1}{2}   
\left. \int_{0}^{1} du  \left[ H^0_{\mu\nu}(x,t,s)
 \right]\right|_{A\rightarrow uA}, \\
\theta_\nu(x,t,s) &=& 
\left. \sum_{y}\int_{0}^{1} du \, 
 \omega_{\nu,\mu\rho}(x,y,t,s)\right|_{A\rightarrow uA}
 \hspace{-8mm}\partial_\rho A_\mu(y,t,s) ,
\end{eqnarray}
we obtain
\begin{equation}
\label{eq:step-one}
q(z)
=\alpha(z) + 
\phi_{\mu\nu}(z-\hat\nu-\hat\mu){F}_{\mu\nu}(z-\hat\nu-\hat\mu)
 + \partial_\nu^* \theta_\nu(z).
\end{equation}

From the topological property of the 4+2 topological field,
\begin{eqnarray}
 0 &=& \int dt ds \,  \, a^4 \sum_{x} \delta q(x,t,s) \nonumber\\
   &=& \int ds dt \, a^4 \sum_{x} \, 
\frac{1}{2} \, 
\phi_{\mu\nu}(z-\hat\nu-\hat\mu)
%\check{F}_{\mu\nu}(z-\hat\nu-\hat\mu)
%\left\{H^0_{\mu\nu}(x-\hat\nu-\hat\mu,t,s)
   \partial_\nu\delta A_\mu (z-\hat\nu-\hat\mu)
%\right\}
.
\end{eqnarray}
By the integration by parts, we obtain
\begin{equation}
\label{eq:div-phi}
\partial_\nu^* 
%\phi_{\mu\nu}(z-\hat\mu)
\phi_{\mu\nu}(z)
=0 .
\end{equation}

\subsubsection{Step two}

Next we examine the differential operator which is obtained
from the variation of $\phi_{\mu\nu}(x,t,s)$.
\begin{equation}
 \delta\phi_{\mu\nu}(z)  
= 
\sum_{y} 
\sum_{m,n =0,1} 
 \frac{\partial \phi_{\mu\nu}(z)}
 {\partial [\partial_s^m\partial_t^n A_\lambda(y,t,s)]} \, 
 \partial_s^m\partial_t^n
  \delta A_\lambda (y,t,s) .
\end{equation}
We denote the differential operator in the above expression
as $L_{\mu\nu\lambda}(x,y,t,s)$.
\begin{equation}
L_{\mu\nu\lambda}(x,y,t,s)
=
\sum_{m,n =0,1} 
 \frac{\partial \phi_{\mu\nu}(z)}
 {\partial [\partial_s^m\partial_t^n A_\lambda(y,t,s)]} \, 
 \partial_s^m\partial_t^n .
\end{equation}
$\phi_{\mu\nu}(x,t,s)$ itself can be expressed with 
$L_{\mu\nu\lambda}(x,y,t,s)$ as 
\begin{gather}
 \phi_{\mu\nu}(z)
= \beta_{\mu\nu}(z)
 +\int_{0}^{1} du \left. \sum_{y}
 L_{\mu\nu,\lambda}(x,y,t,s)\right|_{A\rightarrow u A}
 \hspace{-8mm}A_\lambda (y,t,s).
\end{gather}

It follows from the property Eq.~(\ref{eq:div-phi}) that 
\begin{equation}
 \partial_\mu^*\delta\phi_{\mu\nu}(z)
=\delta \left[\partial_\mu^*\phi_{\mu\nu}(z)\right]=0,
\end{equation}
which implies 
\begin{equation}
 \partial_\mu^* L_{\mu\nu,\lambda}(x,y,t,s)=0.
\end{equation}
and in turn implies
\begin{equation}
  \partial_\mu^\ast \beta_{\mu\nu}(z) = 0 .
\end{equation}
From the gauge invariance of $\phi_{\mu\nu}(z)$,
we have
\begin{equation}
 L_{\mu\nu,\lambda}(x,y,t,s) \partial_\lambda =0.
\end{equation}
Then following the similar argument as the first step, we obtain
\begin{eqnarray}
\label{eq:step-two}
\phi_{\mu\nu}(z)
&=& 
\beta_{\mu\nu}(z) 
\nonumber\\  
&& +
\eta_{\mu\nu\lambda\rho}(z-\hat\lambda-\hat\rho)
 {F}_{\lambda\rho}(z-\hat\lambda-\hat\rho)
\nonumber\\  
&&  + \partial_\rho^* \theta_{\mu\nu\rho}(z),
\end{eqnarray}
and
\begin{equation}
 \partial_\rho^* \eta_{\mu\nu\lambda\rho}(z)=0, \qquad
 \partial_\mu^\ast \beta_{\mu\nu}(z) = 0 .
\end{equation}

\subsubsection{Step three}

Finally, we examine the differential operator which
is obtained from the variation of $\eta_{\mu\nu\lambda\rho}$.
In the course, we encounter the operator of zero degree
$H^0_{\mu\nu\lambda\rho\sigma\tau} : \Omega_0\rightarrow\Omega_6$,
which satisfies the condition
\begin{equation}
 \partial_\tau^* H^0_{\mu\nu\lambda\rho\sigma\tau}(x,t,s)=0.
\end{equation}
$H^0$ is the six form and it may be written with the 
totally antisymmetric tensor $\epsilon_{\mu\nu\lambda\rho\sigma\tau}$
as
\begin{equation}
H^0_{\mu\nu\lambda\rho\sigma\tau}(x,t,s) 
= \delta(x,t,s) \, \, \epsilon_{\mu\nu\lambda\rho\sigma\tau}.
\end{equation}
Then the above condition implies that $\delta$ does not
depend on $(x,t,s)$ and is a constant.

Following the similar argument as the first and second steps, 
we obtain
\begin{eqnarray}
\label{eq:step-three}
\eta_{\mu\nu\lambda\rho}(z)
&=& \gamma_{\mu\nu\lambda\rho}(z) \nonumber\\
&& + \delta \, \epsilon_{\mu\nu\lambda\rho\sigma\tau} \,
 {F}_{\sigma\tau}
 (z-\hat\sigma-\hat\tau)\nonumber\\
&&+ \partial_\sigma^* \theta_{\mu\nu\lambda\rho\sigma}(z),
\end{eqnarray}
and
\begin{equation}
  \partial_\mu^\ast \gamma_{\mu\nu\lambda\rho}(z) = 0,
\end{equation}
where $\epsilon_{\mu\nu\lambda\rho\sigma\tau}$ 
is the totally antisymmetric tensor and $\delta$ is a constant.

Combining these three results and using the Bianchi identity,
\begin{equation}
 \partial^*_{\left[\rho\right.}F_{\left.\mu\nu\right]}
(x-\hat \mu-\hat \nu)
 =
 \partial_{\left[\rho\right.}F_{\left.\mu\nu\right]}
(x-\hat \mu-\hat \nu-\hat \rho)
 = 0,
\end{equation}
%of $F_{\mu\nu}(z)$, 
we finally obtain 
Eq.~(\ref{eq:classification-q}) and complete the proof of the lemma.

In Eq.~(\ref{eq:classification-q}), there is a difference in the 
shifts of the lattice indices from the result obtained in 
\cite{topology-and-axial-anomaly,fujiwara-suzuki-wu}. This difference, 
however, can be shown to be a total divergence.\footnote{
In order to show this, 
we can use the following identity in 
Eqs.~(\ref{eq:step-one}), (\ref{eq:step-two}), 
(\ref{eq:step-three}).
\begin{eqnarray}
&&f_{\mu\nu}(z) g_{\mu\nu}(z) 
-
f_{\mu\nu}(z-\hat \mu - \hat \nu)g_{\mu\nu}(z-\hat \mu - \hat \nu)
\nonumber\\
&=&  
f_{\mu\nu}(z) g_{\mu\nu}(z)  
- f_{\mu\nu}(z-\hat \mu ) g_{\mu\nu}(z-\hat \mu )
\nonumber\\
&& \qquad \qquad \qquad 
+ f_{\mu\nu}(z-\hat \mu ) g_{\mu\nu}(z-\hat \mu )
- f_{\mu\nu}(z-\hat \mu - \hat \nu) g_{\mu\nu}(z-\hat \mu - \hat \nu)
\nonumber\\
&=& 
\partial_i^\ast 
\left( 
  f_{i \nu}(z) g_{i \nu}(z)  
+ f_{\mu i}(z-\hat \mu ) g_{\mu i}(z-\hat \mu )
\right). \nonumber
\end{eqnarray}
}

\section{Exact cancellations of gauge anomalies in \\
SU(2)$_L$ $\times$ U(1)$_Y$ Electroweak theory}
\label{sec:exact-cancellations}
\reseteqnum

The result Eq.~(\ref{eq:classification-q})
concerning the dependence on the 
the admissible U(1)$_Y$ gauge field of the 4+2 dimensional 
topological field may be written symbolically in following form.
\begin{equation}
 q(z;A_\mu,U_\mu) 
= \alpha(z;U_\mu) + \beta(z;U_\mu) {F} + \gamma(z;U_\mu) {F}^2 
         + \delta {F}^3 + d^* k(z;U_\mu,A_\mu)  .
\end{equation}
As we discussed in section~\ref{sec:q-for-electroweak-theory}, 
the 4+2 dimensional topological field for 
SU(2)$_L$ $\times$ U(1)$_Y$ electroweak theory has the properties
\begin{equation}
 q\left(z; A_\mu(z),U_\mu(z)  \right)
= - q\left(z; -A_\mu(z),U_\mu(z) \right)
\end{equation}
and
\begin{equation}
 q\left(z; 0, U_\mu(z) \right)  = 0.
\end{equation}
From these two conditions, we infer 
\begin{equation}
  \alpha(z;U_\mu) = 0.
\end{equation}
and
\begin{equation}
  \gamma(z;U_\mu) {F}^2 
= - d^* \, \frac{1}{2}\left( k(z;U_\mu,-A_\mu) + k(z;U_\mu,A_\mu) \right).
\end{equation}
Therefore, the 4+2 topological field turns out to have 
the following structure
\begin{equation}
 q(z;A_\mu,U_\mu) 
= \beta(z;U_\mu) {F} + \delta {F}^3 
+ d^* 
\, \frac{1}{2}\left( k(z;U_\mu,A_\mu) - k(z;U_\mu,-A_\mu) \right).
\end{equation}
We note that $\beta(z;U_\mu)$ is a gauge-invariant local 
functional of SU(2)$_L$ gauge field, while $\delta$ is a constant 

Now we recall the anomaly cancellation conditions for
the electroweak theory which is given in terms of U(1)$_Y$
hyper-charges. 
Because of the cubic condition
\begin{equation}
 \sum_L Y^3 - \sum_R Y^3 =0 ,
\end{equation}
the term $\delta {F}^3$ vanishes identically, if 
all the contributions from the fermions are summed up. 
On the other hand, because of the linear conditions
\begin{equation}
\sum_{\mbox{doublet(L)}} Y =0. 
\end{equation}
and
\begin{equation}
\sum_{\mbox{singlet(R)}} Y =0 .
\end{equation}
the term $\beta(z;U_\mu) {F}$, which represents
the gauge anomaly of the mixed type, also vanishes identically,
in each sectors of doublets and singlets.
Thus we can see that the 4+2 dimensional topological field
for the electroweak theory is indeed in the trivial cohomology class.
\begin{equation}
 q(z;A_\mu,U_\mu) 
= \partial^*_\mu \, 
\frac{1}{2}\left( k_\mu(z;U_\mu,A_\mu) - k_\mu(z;U_\mu,-A_\mu) \right).
\end{equation}

\section{Summary and discussion}
\label{sec:discussion}
\reseteqnum

We have shown the exact cancellations of gauge anomalies of 
the SU(2)$_L$ $\times$ U(1)$_Y$ electroweak theory on the lattice,
which is formulated based on the lattice Dirac operator 
satisfying the Ginsparg-Wilson relation.
Our approach is to consider the cohomological classification 
of the 4+2-dimensional topological field proposed by L\"uscher
for SU(2)$_L$ $\times$ U(1)$_Y$ electroweak theory.
Using the Poincar\'e lemma in 4+2 dimensions, 
we have determined the dependence on the admissible 
abelian gauge field of U(1)$_Y$ through topological argument, 
{\it with SU(2)$_L$ gauge field fixed as background} 
(cf. \cite{topology-and-axial-anomaly,fujiwara-suzuki-wu}).
This turned out to be sufficient to show
the exact cancellations of the gauge anomalies:
using the pseudo reality of 
SU(2)$_L$ and the anomaly cancellation conditions in the electroweak 
theory given in terms of the hyper-charges of U(1)$_Y$,
we have shown the exact cancellation of the gauge anomaly 
of the mixed type {SU(2)$_L$}$^2$ $\times$ U(1)$_Y$
at finite lattice spacing, as well as {U(1)$_Y$}$^3$.

As to the question of the cohomological classification
of the 4+2 dimensional topological field for the 
electroweak theory, we may also invoke the elegant method based on 
the non-commutative differential calculus and BRST cohomology
in order to explore the structure of the 4+2 dimensional 
topological field \cite{fujiwara-suzuki-wu}.
%
%Another interesting question is where it is possible
%to show the same result in the approach 
%based on the BRST cohomology in four-dimensions.
%

Towards the lattice construction of 
the SU(2)$_L$ $\times$ U(1)$_Y$ electroweak theory,
the next step would be to show the integrability condition given 
in \cite{nonabelian-chiral-gauge-theory}, which assures 
the existence of the functional measure of fermions 
with desired properties.
For this purpose, we need to examine the possible global anomalies
in SU(2)$_L$ $\times$ U(1)$_Y$ electroweak theory.
Global SU(2) anomaly has been examined by Neuberger and
by B\"ar and Campos in detail \cite{neuberger-su2,oliver-isabel}.
It is also desirable to establish the existence of the model
in a finite volume, as in the case of the abelian chiral 
gauge theories \cite{abelian-chiral-gauge-theory}.

In order to extend our result to the whole 
SU(3)$_C$ $\times$ SU(2)$_L$ $\times$ U(1)$_Y$ standard model, 
we need to attack directly the non-abelian nature of the 
gauge anomalies of the mixed type of SU(3)$_C$ $\times$ SU(2)$_L$, 
although there is no corresponding anomaly of this type 
in the continuum theory.
This is also true, of course, when we consider more general
non-ablelian chiral gauge theories.
The recent work by Suzuki 
\cite{BRST-cohomology-4D} based on the BRST cohomology in four-dimensions
could shed lights on this issue.

\section*{Acknowledgments}

The authors are grateful to A.I.~Sanda, K.~Yamawaki, S.~Kitakado,
S.~Uehara, K.~Morita and M.~Harada for discussions and encouragements. 
Y.K. would like to thank S.~Aoki, T.~Onogi, Y.~Taniguchi, 
T.~Izubuchi, K.~Nagai, J.~Noaki, K.~Nagao, N.~Ukita, and H.~So 
for enlightening discussions. The intensive discussions at 
the Summer Institute 99 at Yamanashi, Japan,
were very suggestive and useful to complete this work. 
Y.K. would like to thank M.~L\"uscher, E.~Seiler and P.~Weisz
for their kind hospitality at the Ringberg workshop.
The authors would like to thank M.~L\"uscher and H.~Suzuki 
for valuable comments. 
Y.K. is supported in part by Grant-in-Aid 
for Scientific Research of Ministry of Education (\#10740116).

\end{document}